\pdfoutput=1

\documentclass[11pt]{article}

\usepackage[final]{acl}

\usepackage[T1]{fontenc}
\usepackage[utf8]{inputenc}  
\usepackage{textcomp}        
\usepackage{tgtermes}
\usepackage{latexsym}
\usepackage{tabularx}
\usepackage{float}
\usepackage{amsmath}
\usepackage{bbm}
\usepackage{booktabs}
\usepackage{framed}
\usepackage{mdframed}
\usepackage{enumitem}
\usepackage{hyperref}
\usepackage[T1]{fontenc}
\usepackage{subcaption}
\usepackage[utf8]{inputenc}
\usepackage{placeins}

\usepackage{microtype}

\usepackage{newtxtext,newtxmath}

\usepackage{graphicx}

\title{HyPA-RAG: A Hybrid Parameter Adaptive Retrieval-Augmented Generation System for AI Legal and Policy Applications}

\author{\textbf{Rishi Kalra\textsuperscript{1,2}},
 \textbf{Zekun Wu\textsuperscript{1,2}\thanks{\textbf{Corresponding author}}},
 \textbf{Ayesha Gulley\textsuperscript{1}},
 \textbf{Airlie Hilliard\textsuperscript{1}},\\
  \textbf{Xin Guan\textsuperscript{1}},
\textbf{Adriano Koshiyama\textsuperscript{1}},
\textbf{Philip Treleaven\textsuperscript{2}\footnotemark[1]}\\
\textsuperscript{1}Holistic AI,
\textsuperscript{2}University College London
}

\begin{document}
\maketitle
\begin{abstract}
Large Language Models (LLMs) face limitations in AI legal and policy applications due to outdated knowledge, hallucinations, and poor reasoning in complex contexts. Retrieval-Augmented Generation (RAG) systems address these issues by incorporating external knowledge, but suffer from retrieval errors, ineffective context integration, and high operational costs. This paper presents the Hybrid Parameter-Adaptive RAG (HyPA-RAG) system, designed for the AI legal domain, with NYC Local Law 144 (LL144) as the test case. HyPA-RAG integrates a query complexity classifier for adaptive parameter tuning, a hybrid retrieval approach combining dense, sparse, and knowledge graph methods, and a comprehensive evaluation framework with tailored question types and metrics. Testing on LL144 demonstrates that HyPA-RAG enhances retrieval accuracy, response fidelity, and contextual precision, offering a robust and adaptable solution for high-stakes legal and policy applications.

\end{abstract}

\section{Introduction}

Large Language Models (LLMs) like GPT \citep{Brown2020LanguageMA, Achiam2023GPT4TR}, Gemini \citep{geminiteam2024geminifamilyhighlycapable}, and Llama \citep{Touvron2023LLaMAOA, Touvron2023Llama2O, Dubey2024TheL3} have advanced question answering across domains \citep{Brown2020LanguageMA, Singhal2023TowardsEM, Wu2023BloombergGPTAL}. However, they face challenges in domains like law and policy due to outdated knowledge limited to pre-training data \citep{Yang2023HarnessingTP} and hallucinations, where outputs appear plausible but are factually incorrect \citep{Ji2022SurveyOH, Huang2023ASO}.
Empirical evidence indicates that many AI tools for legal applications overstate their ability to prevent hallucinations \citep{Magesh2024HallucinationFreeAT}. Cases of lawyers penalized for using hallucinated court documents \citep{fortune2023, businessinsider2023} highlight the need for reliable AI systems in legal and policy contexts.

Retrieval-Augmented Generation (RAG) integrates external knowledge into LLMs to address their limitations but faces challenges. These include missing content, where relevant documents are not retrieved; context limitations, where retrieved documents are poorly integrated into responses; and extraction failures due to noise or conflicting data \citep{Barnett2024SevenFP}. Advanced techniques like query rewriters and LLM-based quality checks improve quality but increase token usage and costs.

\begin{figure*}[h]
  \includegraphics[width=1\linewidth]{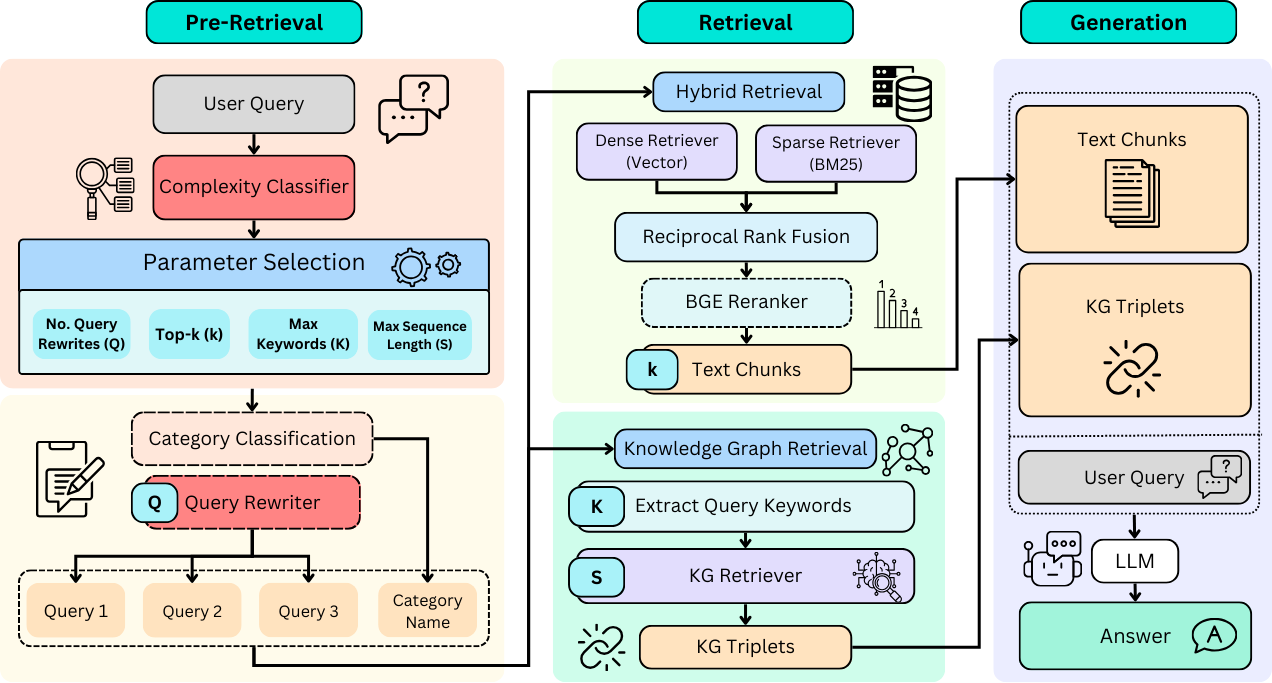}
  \caption{Hybrid Parameter Adaptive RAG (HyPA-RAG) System Diagram}
  \label{fig:system-diagram}
\end{figure*}

This research presents the Hybrid Parameter-Adaptive RAG (HyPA-RAG) system to address RAG challenges in AI policy, using NYC Local Law 144 as a test corpus. HyPA-RAG includes adaptive parameter selection with a query complexity classifier to reduce token usage, a hybrid retrieval system combining dense, sparse, and knowledge graph methods to improve accuracy, and an evaluation framework with a gold dataset, custom question types, and RAG-specific metrics. These components address common RAG failures and enhance AI applications in legal and policy domains.


\section{Background and Related Work}

Recent LLM advancements have influenced law and policy, where complex language and large text volumes are common \citep{BlairStanek2023CanGP, Choi2023ChatGPTGT, Hargreaves2023WordsAF}. LLMs have been applied to legal judgment prediction, document drafting, and contract analysis, improving efficiency and accuracy \citep{shui2023, sun2023, savelka2023}. Techniques like fine-tuning, retrieval augmentation, prompt engineering, and agentic methods have further enhanced performance in summarization, drafting, and interpretation \citep{Trautmann2022LegalPE, Cui2023ChatlawAM}.


RAG enhances language models by integrating external knowledge through indexing, retrieval, and generation, using sparse (e.g., BM25) and dense (e.g., vector) techniques with neural embeddings to improve response specificity, accuracy, and grounding \citep{Lewis2020RetrievalAugmentedGF, Gao2023RetrievalAugmentedGF, SprckJones2021ASI, Robertson2009ThePR, Devlin2019BERTPO, Liu2019RoBERTaAR}. To overcome naive RAG's limitations, such as poor context and retrieval errors, advanced methods like hybrid retrieval, query rewriters, and rerankers have been developed \citep{muennighoff2022mteb, Ding2024HybridLC, bge_embedding}. Hybrid retrieval combines BM25 with semantic embeddings for better keyword matching and contextual understanding \citep{Luo2023StudyHybridRetrievers, Ram2022LearningPassages, Arivazhagan2023HybridHierarchical}, while knowledge graph retrieval and composed retrievers improve accuracy and comprehensiveness \citep{Rackauckas2024RAGFusionAN, Sanmartin2024KGRAGBT, Edge2024FromLT}. Recently, RAG systems have advanced from basic retrieval to dynamic methods involving multi-source integration and domain adaptation \citep{Gao2023RetrievalAugmentedGF, Ji2022SurveyOH}. Innovations like Self-RAG and KG-RAG improve response quality and minimize hallucinations through adaptive retrieval and knowledge graphs \citep{asai2023selfrag, Sanmartin2024KGRAGBT}. Frameworks for evaluating RAG systems include Ragas, which uses reference-free metrics like faithfulness and relevancy \citep{Shahul2023RAGAsAE}, Giskard, which leverages synthetic QA datasets \citep{giskard2023}, and ARES, which employs prediction-powered inference with LLM judges for precise evaluation \citep{giskard2023, SaadFalcon2023ARESAA}.

\section{System Design}
The Hybrid Parameter-Adaptive RAG (HyPA-RAG) system, shown in Figure \ref{fig:system-diagram}, integrates vector-based text chunks and a knowledge graph of entities and relationships to improve retrieval accuracy. It employs a hybrid retrieval process that combines sparse (BM25) and dense (vector) methods to retrieve an initial top-$k$ set of results, refined using reciprocal rank fusion based on predefined parameter mappings. A knowledge graph (KG) retriever dynamically adjusts retrieval depth and keyword selection based on query complexity, retrieving relevant triplets. Results are combined with the KG results appending it to the retrieved chunks to generate an final set of $k$ chunks. Optional components include a query rewriter to enhance retrieval with reformulated queries and a reranker for further refining chunk ranking. De-duplicated rewritten query results are integrated into the final set, which, along with knowledge graph triplets, is processed within the LLM's context window for precise, contextually relevant responses. The framework has two variations: Parameter-Adaptive (PA) RAG, which excludes knowledge graph retrieval, and Hybrid Parameter-Adaptive (HyPA) RAG, which incorporates it.

\section{AI Legal and Policy Corpus}

Local Law 144 (LL144) of 2021, enacted by New York City’s Department of Consumer and Worker Protection (DCWP), regulates automated employment decision tools (AEDTs). This study uses a 15-page version of LL144, combining the original law with DCWP enforcement rules. As an early AI-specific law, LL144 is included in GPT-4 and GPT-4o training data, verified via manual prompting, and serves as a baseline in this research. The complexity of LL144 motivates our system's design for several reasons: (1) it requires multi-step reasoning and concept linking due to its mix of qualitative and quantitative requirements—definitions, procedural guidelines, and compliance metrics—that semantic similarity alone cannot capture, addressed through our knowledge graph; (2) seemingly simple queries can be ambiguous or require multiple information chunks, making a query rewriter and classifier necessary; and (3) while not specific to our adaptive classifier, the evolving nature of AI laws limits the effectiveness of static pre-training, making retrieval-augmented systems better suited to handle frequent updates. These factors go beyond what standard LLMs and basic RAG systems can manage, justifying the need for our approach.

\section{Performance Evaluation}
The evaluation process starts by generating custom questions tailored to AI policy and legal question-answering, then introduces and verifies evaluation metrics (see evaluation section of Figure \ref{fig:overall-rag-workflow} in Appendix \ref{appendix:workflow-diagram}). \textbf{For reproducibility, the LLM temperature is set to zero for consistent responses and all other parameters are set to defaults.}

\subsection{Dataset Generation}
We created a "gold standard" evaluation set to assess system performance, leveraging GPT-3.5-Turbo and Giskard \citep{giskard2023} for efficient question generation. The dataset includes various question types, such as 'simple', 'complex', 'situational', and novel types like 'comparative', 'complex situational', 'vague', and 'rule-conclusion' (inspired by LegalBench \citep{Guha2023LegalBenchAC}). These questions test multi-context retrieval, user-specific contexts, query interpretation, and legal reasoning. Generated questions were deduplicated and refined through expert review to ensure accuracy and completeness, using the criteria outlined in Table \ref{tab:evaluation_criteria} in Appendix \ref{subsec:annotation-criteria}.

\subsection{Evaluation Metrics}

To evaluate our RAG system, we utilise RAGAS metrics \citep{shahul2023ragas} based on the LLM-as-a-judge approach \citep{Zheng2023JudgingLW}, including Faithfulness, Answer Relevancy, Context Precision, Context Recall, and an adapted Correctness metric.


\textbf{Faithfulness} evaluates the factual consistency between the generated answer and the context, defined as $\text{Faithfulness Score} = \frac{|C_{\text{inferred}}|}{|C_{\text{total}}|}$, where $C_{\text{inferred}}$ is the number of claims inferred from the context, and $C_{\text{total}}$ is the total claims in the answer.

\textbf{Answer Relevancy} measures the alignment between the generated answer and the original question, calculated as the mean cosine similarity between the original question and generated questions from the answer: $\text{Answer Relevancy} = \frac{1}{N} \sum_{i=1}^{N} \frac{E_{g_i} \cdot E_{o}}{\Vert E_{g_i} \Vert \Vert E_{o} \Vert}$, where $E_{g_i}$ and $E_{o}$ are embeddings of the generated and original questions.


\textbf{Context Recall} measures the proportion of ground truth claims covered by the retrieved context, defined as $\text{Context Recall} = \frac{|C_{\text{attr}}|}{|C_{\text{GT}}|}$, where $C_{\text{attr}}$ is the number of ground truth claims attributed to the context, and $C_{\text{GT}}$ is the total number of ground truth claims.



\textbf{Context Precision} evaluates whether relevant items are ranked higher within the context, defined as $\text{Context Precision} = \frac{\sum_{k=1}^{K} (\text{$P_k$} \times v_k)}{|R_k|}$. Here, $\text{$P_k$} = \frac{TP_k}{TP_k + FP_k}$ is the precision at rank $k$, $v_k$ is the relevance indicator, $|R_k|$ is the total relevant items in the top $K$, $TP_k$ represents true positives, and $FP_k$ false positives.


\subsection{Correctness Evaluation}

We assess correctness using a refined metric to address the limitations of Giskard's binary classification, which fails to account for partially correct answers or minor variations. Our adapted metric, \textbf{Absolute Correctness}, based on LLamaIndex \citep{llamaindex}, uses a 1 to 5 scale: 1 indicates an incorrect answer, 3 denotes partial correctness, and 5 signifies full correctness. For binary evaluation, we use a high threshold of 4, reflecting our low tolerance for inaccuracies. The \textbf{Correctness Score} is computed as the average of these binary outcomes across all responses:
$\text{Correctness Score} = \frac{1}{N} \sum_{i=1}^{N} \mathbbm{1}(S_i \geq 4),$
where $S_i$ represents the absolute correctness score of the $i$th response, $\mathbbm{1}(S_i \geq 4)$ is an indicator function that is 1 if $S_i \geq 4$ and 0 otherwise, and $N$ is the total number of responses.

\begin{figure}[t]
  \includegraphics[width=\columnwidth]{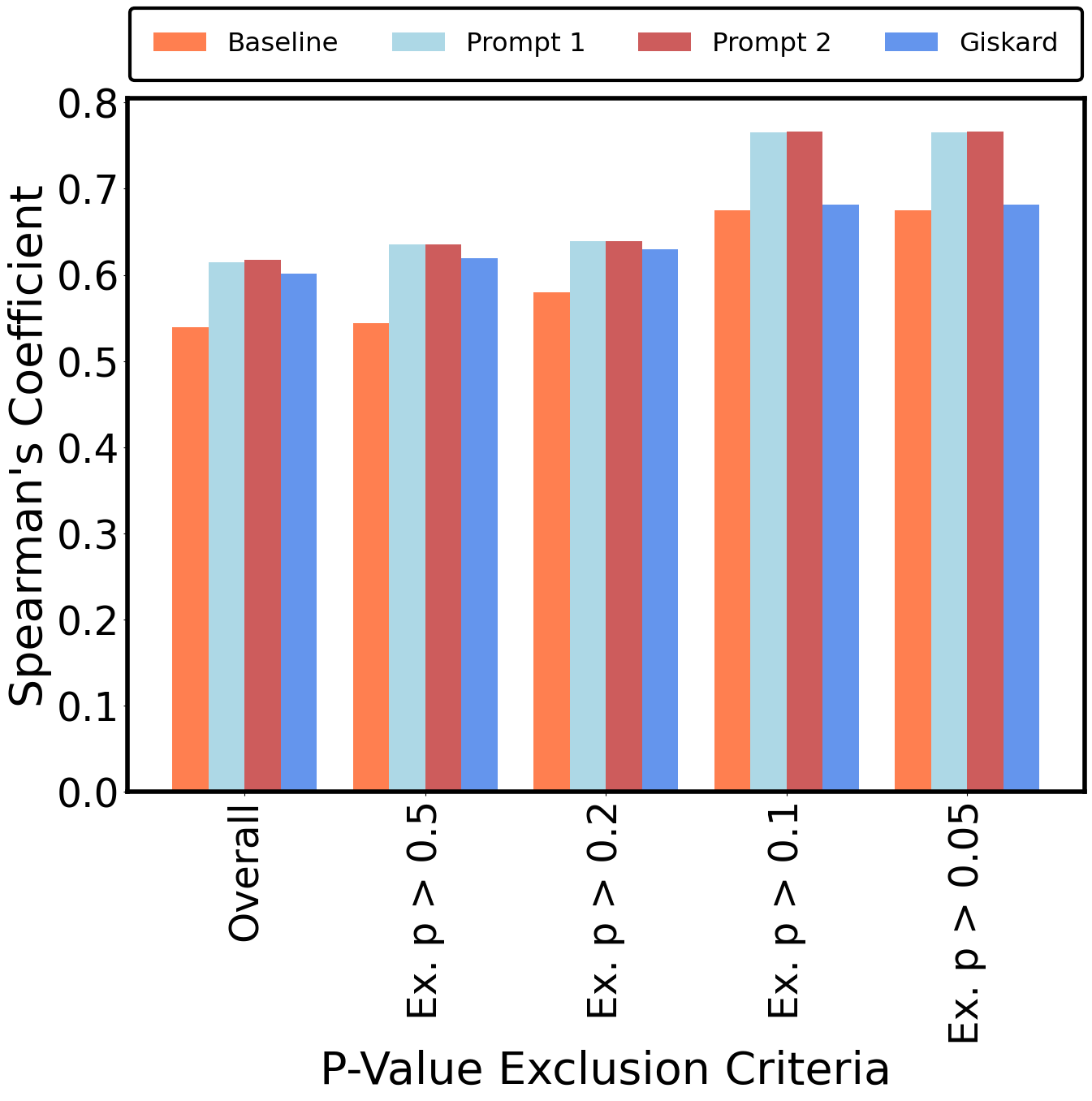}
  \caption{\textbf{Spearman Coefficient Comparison}, showing the correlation between model performance and human evaluation.}
  \label{fig:spearman_overall}
\end{figure}

The Spearman coefficient (Figure \ref{fig:spearman_overall}) shows how our prompted LLM correctness judge aligns with human judgment. Prompts 1 and 2 (Appendix \ref{appendix:correctness-eval-prompts}) employ different methods: the baseline prompt provides general scoring guidelines, Prompt 1 offers detailed refinements, and Prompt 2 includes one-shot examples and edge cases.

Additional metrics, including macro precision, recall, F1 score, and percentage agreement with human labels, are shown in Figure \ref{fig:correctness_classification_metrics} (Appendix \ref{appendix:correctness-eval-results}). A detailed breakdown of the Spearman coefficient metrics is provided in Figure \ref{fig:correctness_spearman_metrics} (Appendix \ref{appendix:correctness-eval-results}).

\section{Chunking Method}

We evaluate three chunking techniques: sentence-level, semantic, and pattern-based chunking.

Sentence-level chunking splits text at sentence boundaries, adhering to token limits and overlap constraints. Semantic chunking uses cosine similarity to set a dissimilarity threshold for splitting and includes a buffer size to define the minimum number of sentences before a split. Pattern-based chunking employs a custom delimiter based on text structure; for LL144, this is "\verb|\n|\textsection". 

\begin{figure}[t]
  \includegraphics[width=\columnwidth]{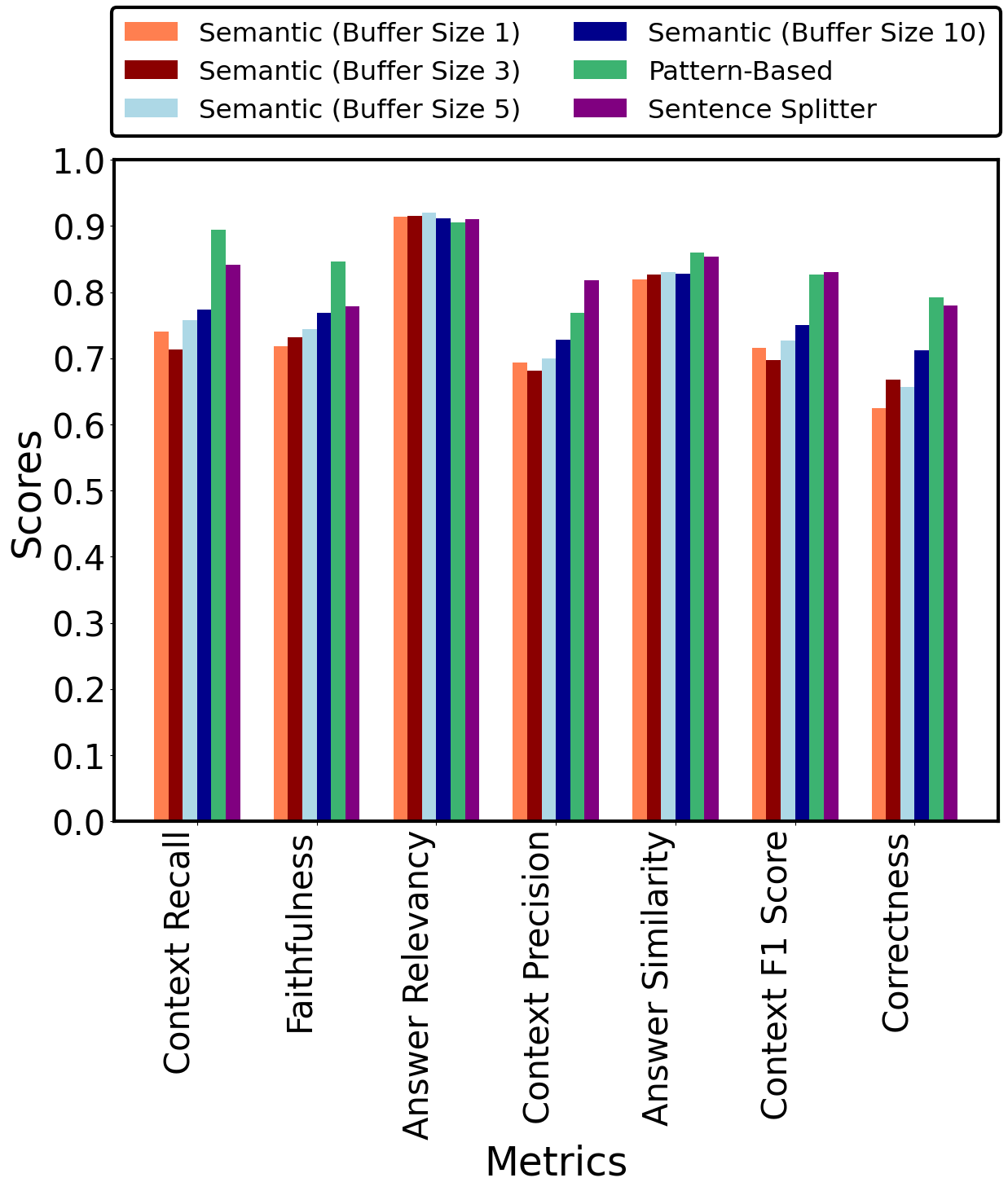}
  \caption{RAG Evaluation Metrics for Sentence-Level, Semantic, and Pattern-Based Chunking Methods}
  \label{fig:chunking_results}
\end{figure}

Figure \ref{fig:chunking_results} shows that pattern-based chunking achieves the highest context recall (0.9046), faithfulness (0.8430), answer similarity (0.8621), and correctness (0.7918) scores. Sentence-level chunking, however, yields the highest context precision and F1 scores. Semantic chunking performs reasonably well with increased buffer size but generally underperforms compared to the simpler methods. Further hyperparameter tuning may improve its effectiveness. These findings suggest that a corpus-specific delimiter can enhance performance over standard chunking methods.

For subsequent experiments, we adopt sentence-level chunking with a default chunk size of 512 tokens and an overlap of 200 tokens. 

\section{Query Complexity Classifier}

We developed a domain-specific query complexity classifier for adaptive parameter selection, mapping queries to specific hyper-parameters. Unlike Adaptive RAG \citep{jeong-etal-2024-adaptive}, our classifier influences not only the top-$k$ but also knowledge graph and query rewriter parameters. Our analysis of top-$k$ selection indicated different optimal top-$k$ values for various question types, as shown in Figure \ref{fig:eval-metrics-vary-topk} (Appendix \ref{appendix:eval-results-vary-topk}).

\subsection{Training Data}

To train a domain-specific query complexity classifier, we generated a dataset using a GPT-4o model on legal documents. Queries were categorised into three classes based on the number of contexts required: one context (0), two contexts (1), and three or more contexts (2). This classification resulted in varying token counts, keywords, and clauses across classes, which could bias models toward associating these features with complexity. To mitigate this, we applied data augmentation techniques to diversify the dataset.
To enhance robustness, 67\% of the queries were modified. We increased vagueness in 10\% of the questions while preserving their informational content, added random noise words or punctuation to another 10\%, and applied both word and punctuation noise to a further 10\%. Additionally, 5\% of questions had phrases reordered, and another 5\% contained random spelling errors. For label-specific augmentation, 25\% of label 0 queries were made more verbose, and 25\% of label 2 queries were shortened, ensuring they retained the necessary informational content. The augmentation prompts are in Appendix \ref{appendix:classifier-data-aug-prompts}.
\begin{table}
  \centering
  \resizebox{\columnwidth}{!}{%
  \begin{tabular}{llll}
    \hline
    \textbf{Model} & \textbf{Precision} & \textbf{Recall} & \textbf{F1 Score} \\
    \hline
    Random Labels & 0.34 & 0.34 & 0.34 \\
    BART Large ZS & 0.31 & 0.32 & 0.29 \\
    DeBERTa-v3 ZS & 0.39 & 0.39 & 0.38 \\
    LR TF-IDF & 0.84 & 0.84 & 0.84 \\
    SVM TF-IDF & 0.86 & 0.86 & 0.86 \\
    distilBERT Finetuned & 0.90 & 0.90 & 0.90 \\
    \hline
  \end{tabular}
  }
  \caption{3-Class Classification Results}
  \label{tab:classification-results-3class}
\end{table}

\begin{table*}[t]
\centering
\resizebox{\textwidth}{!}{%
\begin{tabular}{p{4.5cm}p{2.5cm}p{2.5cm}p{3.cm}p{2.5cm}}
\toprule
\textbf{Method} & \textbf{Faithfulness} & \textbf{Answer \newline Relevancy} & \textbf{Absolute \newline Correctness (1-5)} & \textbf{Correctness \newline(Threshold=4.0)} \\
\midrule
\multicolumn{5}{l}{\textbf{LLM Only}} \\
\midrule
GPT-3.5-Turbo & 0.2856 & 0.4350 & 2.6952 & 0.1973 \\
GPT-4o-Mini & 0.3463 & 0.6319 & 3.3494 & 0.4572 \\
\midrule
\multicolumn{5}{l}{\textbf{Fixed $k$}} \\
\midrule
$k = 3$ & 0.7748 & 0.7859 & 4.0372 & 0.7546 \\
$k = 5$ & 0.8113 & 0.7836 & 4.0520 & 0.7584 \\
$k = 7$ & 0.8215 & 0.7851 & 4.0520 & 0.7621 \\
$k = 10$ & 0.8480 & 0.7917 & 4.0595 & 0.7658 \\
\midrule
\multicolumn{5}{l}{\textbf{Adaptive}} \\
\midrule
PA: $k, Q$ (2 class) & \textbf{0.9044} & \textbf{0.7910} & \underline{4.2491} & \underline{0.8104} \\
PA: $k, Q$ (3 class) & \underline{0.8971} & 0.7778 & \textbf{4.2528} & \textbf{0.8141} \\
HyPA: $k, Q, K, S$ (2 class) & 0.8328 & \underline{0.7800} & 4.0558 & 0.7770 \\
HyPA: $k, Q, K, S$ (3 class) & 0.8465 & 0.7734 & 4.1338 & 0.7918 \\
\bottomrule
\end{tabular}%
}
\caption{Performance metrics for LLM Only, Fixed $k$, Parameter-Adaptive (PA), and Hybrid Parameter Adaptive (HyPA) RAG implementations for the 2 and 3-class classifier configurations. $k$ is the top-$k$ value, $Q$ the number of query rewrites, $S$ the maximum knowledge graph depth, and $K$ the maximum keywords for knowledge graph retrieval.}
\label{tab:combined-results-metrics}
\end{table*}
\subsection{Model Training}

We employed multiple models as baselines for classification tasks: Random labels, Logistic Regression (LR), Support Vector Machine (SVM), zero-shot classifiers, and a fine-tuned DistilBERT model. The Logistic Regression model used TF-IDF features, with a random state of 5 and 1000 iterations. The SVM model also used TF-IDF features with a linear kernel. Both models were evaluated on binary (2-class) and multi-class (3-class) tasks. Zero-shot classifiers (BART Large ZS and DeBERTa-v3 ZS) were included as additional baselines, mapping "simple question," "complex question," and "overview question" to labels 0, 1, and 2, respectively; for binary classification, only "simple question" (0) and "complex question" (1) were used. The DistilBERT model was fine-tuned with a learning rate of 2e-5, batch size of 32, 10 epochs, and a weight decay of 0.01 to optimize performance and generalization to the validation set.

\subsection{Classifier Results}

Tables \ref{tab:classification-results-3class} and \ref{tab:classification-results-2class} in Appendix \ref{appendix:2-class-classifier-results} summarise the classification results. We compare performance using macro precision, recall and F1 score. The fine-tuned DistilBERT model achieved the highest F1 scores, 0.90 for the 3-class task and 0.92 for the 2-class task, highlighting the benefits of transfer learning and fine-tuning. The SVM (TF-IDF) and Logistic Regression models also performed well, particularly in binary classification, indicating their effectiveness in handling sparse data. Zero-shot classifiers performed lower.

\section{RAG System Architecture}
\subsection{Parameter-Adaptive RAG (PA-RAG)}

The Parameter-Adaptive RAG system integrates our fine-tuned DistilBERT model to classify query complexity and dynamically adjusts retrieval parameters accordingly, as illustrated in Figure \ref{fig:system-diagram}, but excluding the knowledge graph component. The PA-RAG system adaptively selects the number of query rewrites ($Q$) and the top-$k$ value based on the complexity classification, with specific parameter mappings provided in Table \ref{tab:parameter_mappings_part1} in Appendix \ref{appendix:param-mappings-1}. In the 2-class model, simpler queries (label 0) use a top-$k$ of 5 and 3 query rewrites, while more complex queries (label 1) use a top-$k$ of 10 and 5 rewrites. The 3-class model uses a top-$k$ of 7 and 7 rewrites for the most complex queries (label 2).

\subsection{Hybrid Parameter-Adaptive RAG}

Building on the PA-RAG system, the Hybrid Parameter-Adaptive RAG (HyPA-RAG) approach enhances the retrieval stage by addressing issues such as missing content, incomplete answers, and failures of the language model to extract correct answers from retrieved contexts. These challenges often arise from unclear relationships within legal documents, where repeated terms lead to fragmented retrieval results \citep{Barnett2024SevenFP}. Traditional (e.g. dense) retrieval methods may retrieve only partial context, causing missing critical information. To overcome these limitations, this system incorporates a knowledge graph (KG) representation of LL144. Knowledge graphs, structured with entities, relationships, and semantic descriptions, integrate information from multiple data sources \citep{Hogan2020KnowledgeG, Ji2020ASO}, and recent advancements suggest that combining KGs with LLMs can produce more informed outputs using KG triplets as added context.

The HyPA-RAG system uses the architecture outlined in Figure \ref{fig:system-diagram}. The knowledge graph is constructed by extracting triplets (subject, predicate, object) from raw text using GPT-4o. Parameter mappings specific to this implementation, such as the maximum number of keywords per query ($K$) and maximum knowledge sequence length ($S$), are detailed in Table \ref{tab:parameter_mappings_part2}, extending those provided in Table \ref{tab:parameter_mappings_part1}.

\subsection{RAG Results}

Adaptive methods consistently outperform fixed $k$ baselines. PA-RAG $k, Q$ (2 class) achieves the highest faithfulness score of 0.9044, a 0.0564 improvement over the best fixed method ($k = 10$, 0.8480). Similarly, PA $k, Q$ (3 class) achieves 0.8971, surpassing all fixed $k$ methods. For answer relevancy, PA $k, Q$ (2 class) scores 0.7910, nearly matching the best fixed method (0.7917), while PA $k, Q$ (3 class) scores slightly lower at 0.7778. In absolute correctness, PA $k, Q$ (2 class) and $k, Q$ (3 class) achieve 4.2491 and 4.2528, respectively, improving by 0.1896 and 0.1933 over the best fixed method ($k = 10$, 4.0595). Correctness scores further highlight the advantage, with PA $k, Q$ (3 class) scoring 0.8141, 0.0483 higher than the fixed baseline (0.7658). HyPA results are more variable. HyPA $k, Q, K, S$ (2 class) achieves a correctness score of 0.7770, a modest 0.0112 improvement over fixed $k = 7$, indicating potential for further optimization.

\subsection{System Ablation Study}

We evaluate the impact of adaptive parameters, a reranker (bge-reranker-large), and a query rewriter on model performance using PA and HyPA RAG methods with 2-class (Table \ref{tab:ablation-2-class} in Appendix \ref{appendix:2-class-ablation}) and 3-class classifiers (Table \ref{tab:ablation-3-class} in Appendix \ref{appendix:3-class-ablation}).

Adaptive parameters, query rewriting, and reranking significantly influence RAG performance. Varying the top-$k$ chunks alone achieves the highest Answer Relevancy (0.7940), while adapting the top-$k$ and number of query rewrites with a reranker ($k$, $Q$ + reranker) delivers the highest Faithfulness (0.9098) and improves Correctness Score from 0.8141 to 0.8178. Adding a knowledge graph ($k$, $K$, $S$) maintains the same Correctness Score (0.8141) but lowers Absolute Correctness. The HyPA ($k$, $K$, $S$, $Q$ + reranker) setup achieves the highest Correctness Score (0.8402), showing the value of adaptive parameters and reranking in improving correctness.


\section{Overall Results and Discussion}

Our analysis demonstrates that adaptive methods outperform fixed baselines, particularly in faithfulness and answer quality. Adaptive parameters, such as query rewrites and reranking, enhance response accuracy and relevance, though reranking may slightly reduce overall correctness scores, indicating a trade-off between precision and quality. Adding a knowledge graph maintains correctness but introduces complexity, potentially lowering response quality. However, combining adaptive parameters with reranking maximizes correct responses, even if it doesn't achieve the highest scores across all metrics. These findings demonstrate the effect of adaptivity and parameter tuning to balance performance, enabling effective handling of diverse and complex queries. This suggests our system could also apply to other domains where queries demand complex, multi-step reasoning and non-obvious concept relationships. \textbf{Limitations and future work are detailed in Appendix \ref{appendix:future-work-limitations}.}

\section{Ethical Considerations}
The deployment of the Hybrid Parameter-Adaptive RAG (HyPA-RAG) system in AI legal and policy contexts raises critical ethical and societal concerns, particularly regarding the accuracy, reliability, and potential misinterpretation of AI-generated responses. The high-stakes nature of legal information means inaccuracies could have significant consequences, highlighting the necessity for careful evaluation. We emphasize transparency and reproducibility, providing detailed documentation of data generation, retrieval methods, and evaluation metrics to facilitate replication and scrutiny. The environmental impact of NLP models is also a concern. Our system employs adaptive retrieval strategies to optimize computational efficiency, reduce energy consumption, and minimize carbon footprint, promoting sustainable AI development. Our findings enhance the understanding of RAG systems in legal contexts but are intended for research purposes only. HyPA-RAG outputs should not be used for legal advice or decision-making, emphasizing the need for domain expertise and oversight in applying AI to sensitive legal domains.


\bibliography{custom}

\clearpage

\appendix
\onecolumn
\section{Appendix}
\label{sec:appendix}

\subsection{RAG Demonstration User Interface}
\label{appendix:rag-demo-ss}

\begin{figure}[H]
    \centering
    
    \begin{subfigure}[b]{0.8\textwidth}
        \centering
        \includegraphics[width=\linewidth]{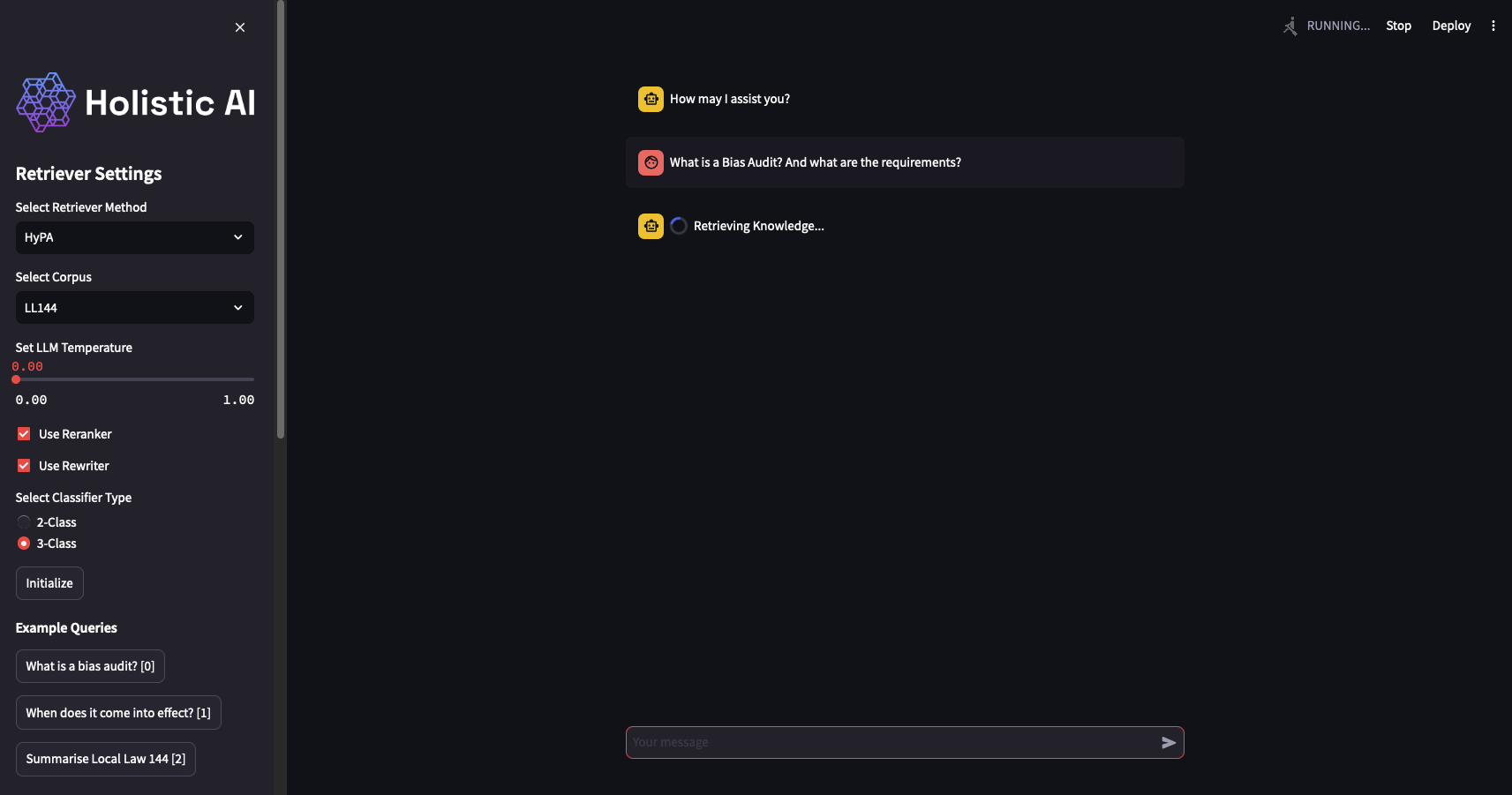}
        \caption{Demo Screenshot: Entering the user query and generating a response.}
        \label{fig:sub-fig1}
    \end{subfigure}
    \hfill
    \begin{subfigure}[b]{0.8\textwidth}
        \centering
        \includegraphics[width=\linewidth]{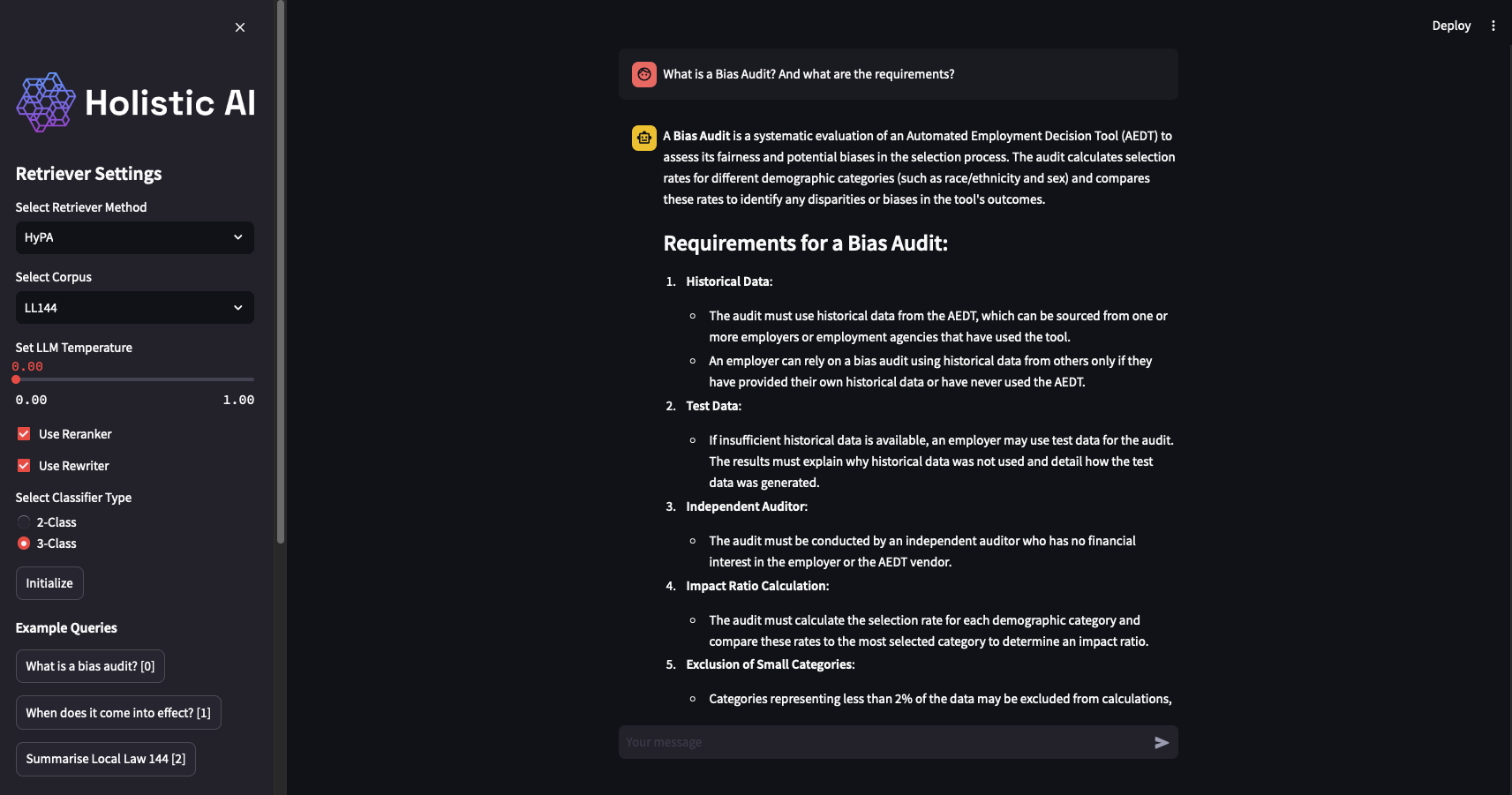}
        \caption{Demo Screenshot: The generated response.}
        \label{fig:sub-fig2}
    \end{subfigure}
    \hfill
    \begin{subfigure}[b]{0.8\textwidth}
        \centering
        \includegraphics[width=\linewidth]{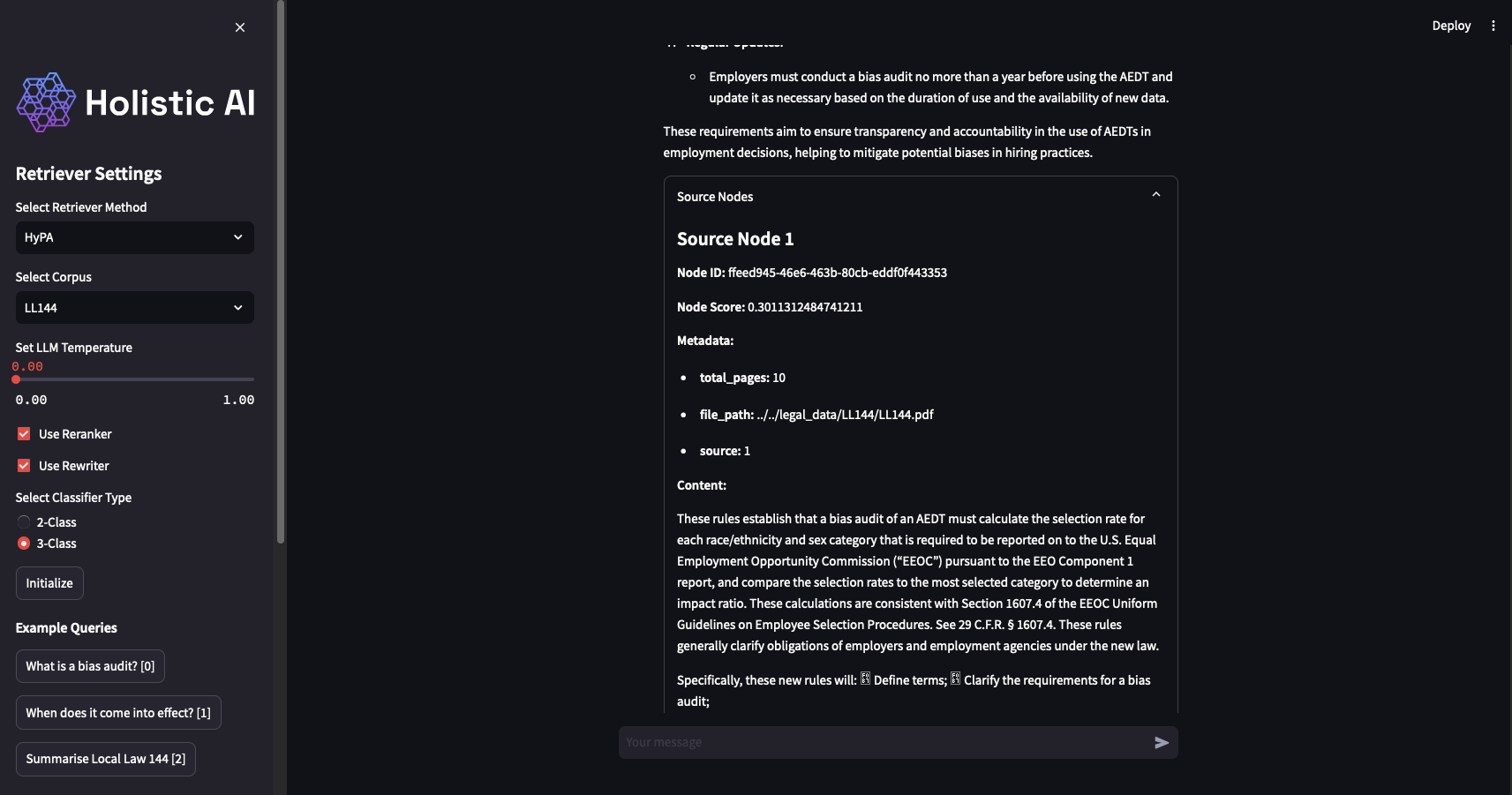}
        \caption{Demo Screenshot: Information on retrieved node metadata and content.}
        \label{fig:sub-fig3}
    \end{subfigure}
    
    \caption{Demo screenshots showing each key stage of the user experience.}
    \label{fig:hypa-rag-demo}
\end{figure}
\clearpage
\subsection{Overall Workflow Diagram}
\label{appendix:workflow-diagram}

\begin{figure*}[h]
    \centering
\includegraphics[width=14cm]{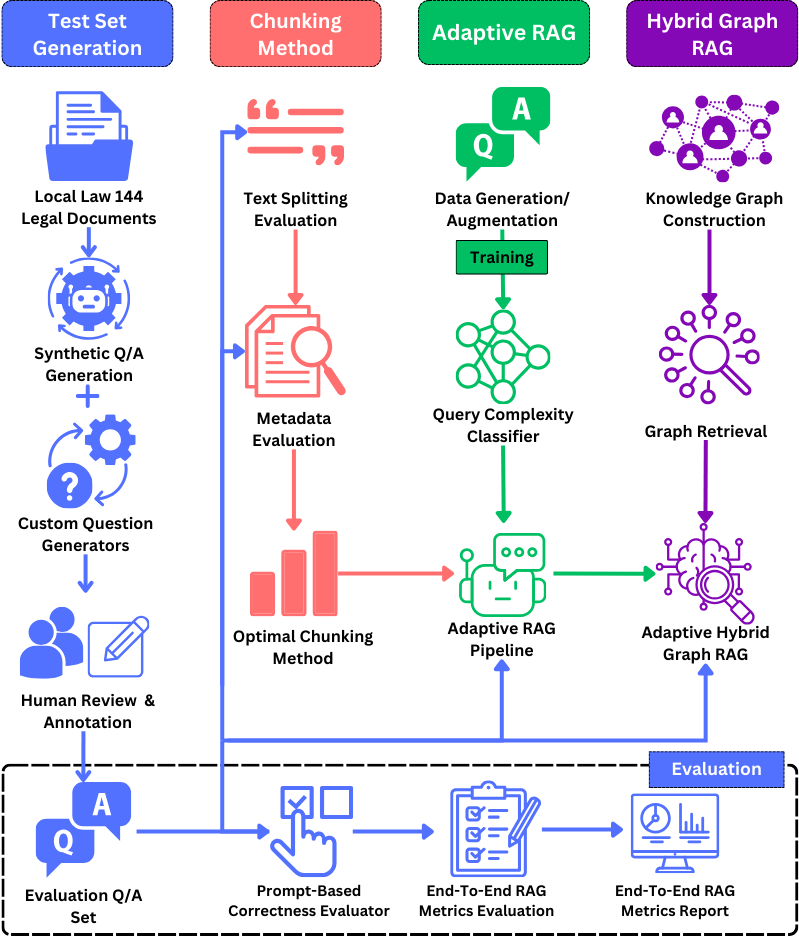}
  \caption {Overall RAG Development Workflow Diagram}
  \label{fig:overall-rag-workflow}
\end{figure*}

\clearpage
\subsection{Question Types}
\label{subsec:question-types}

\begin{table}[H]
  \centering
  \begin{tabular}{p{2.4cm}p{4cm}p{6cm}p{2cm}}
    \hline
    \textbf{Question \newline Type} & \textbf{Description} & \textbf{Example \newline Question} & \textbf{Target RAG \newline Components} \\
    \hline
    Simple & Requires retrieval of one concept from the context & What is a bias audit? & Generator, Retriever, Router \\
    \hline
    Complex & More detailed and requires more specific retrieval & What is the purpose of a bias audit for automated employment decision tools? & Generator, Retriever \\
    \hline
    Distracting & Includes an irrelevant distracting element & Italy is beautiful but what is a bias audit? & Generator, Retriever, Rewriter \\
    \hline
    Situational & Includes user context to produce relevant answers & As an employer, what information do I need to provide before using an AEDT? & Generator \\
    \hline
    Double & Two distinct parts to evaluate query rewriter & What are the requirements for a bias audit of an AEDT and what changes were made in the second version of the proposed rules? & Generator, Rewriter \\
    \hline
    Conversational & Part of a conversation with context provided in a previous message & (1) I would like to know about bias audits. (2) What is it? & Rewriter \\
    \hline
    Complex situational & Introduces further context and one or more follow-up questions within the same message & In case I need to recover a civil penalty, what are the specific agencies within the office of administrative trials and hearings where the proceeding can be returned to? Also, are there other courts where such a proceeding can be initiated? & Generator \\
    \hline
    Out of scope & Non-answerable question that should be rejected & Who developed the AEDT software? & Generator, Prompt \\
    \hline
    Vague & A vague question that lacks complete information to answer fully & What calculations are required? & Generator, Rewriter \\
    \hline  
    Comparative & Encourages comparison and identifying relationships & What are the differences and similarities between 'selection rate' and 'scoring rate', and how do they relate to each other? & Generator, Rewriter \\
    \hline
    Rule conclusion & Provides a scenario, requiring a legal conclusion & An employer uses an AEDT to screen candidates for a job opening. Is the selection rate calculated based on the number of candidates who applied for the position or the number of candidates who were screened by the AEDT? & Generator, Rewriter \\
    \hline
  \end{tabular}
  \caption{\label{tab:questions} Question types and their descriptions with targeted RAG components.}
\end{table}

\FloatBarrier
\clearpage

\subsection{Evaluation Results for Varied Top-$k$}
\label{appendix:eval-results-vary-topk}
\begin{figure*}[h]
  \includegraphics[width=0.32\linewidth]{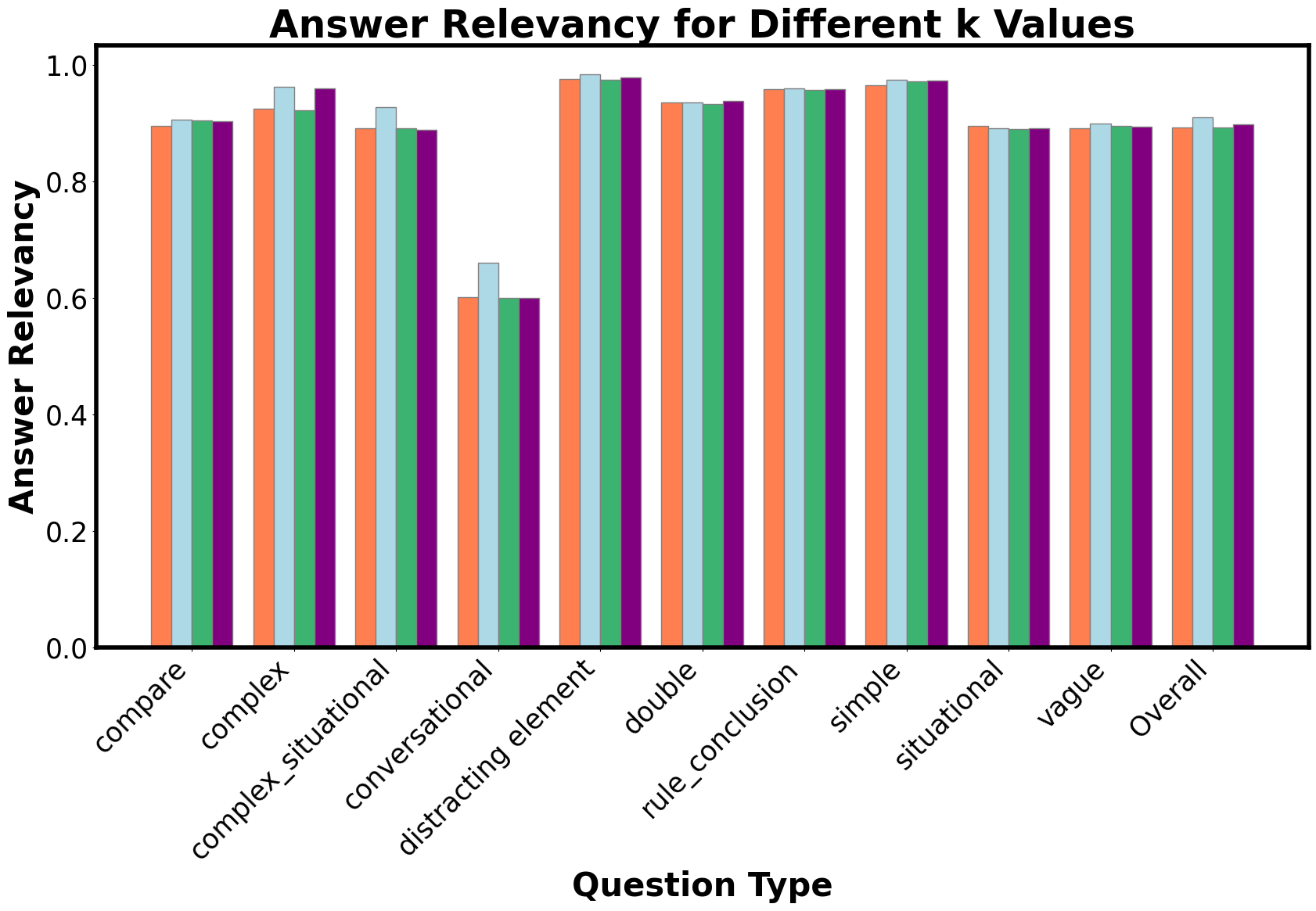} \hfill
  \includegraphics[width=0.32\linewidth]{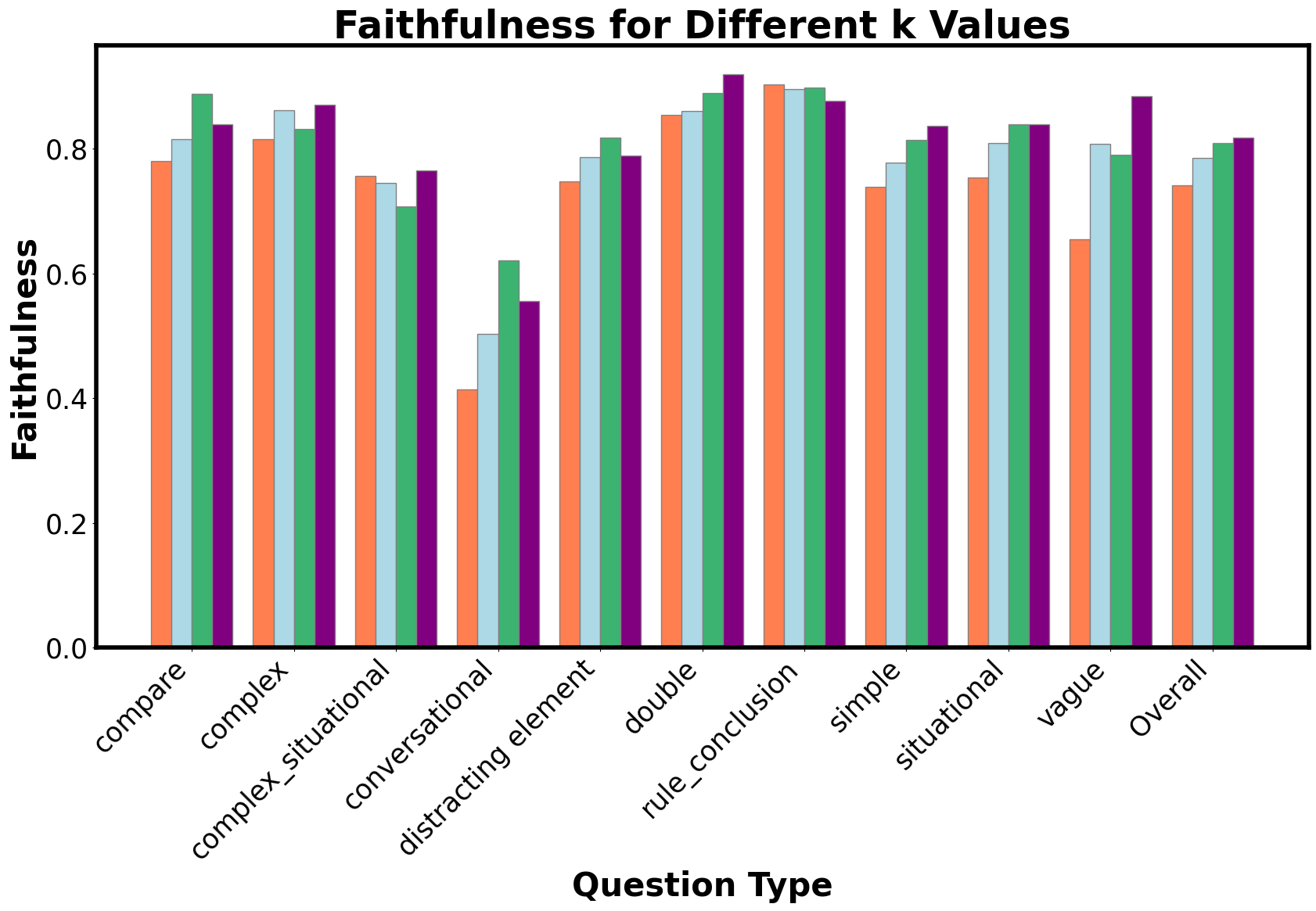} \hfill
  \includegraphics[width=0.32\linewidth]{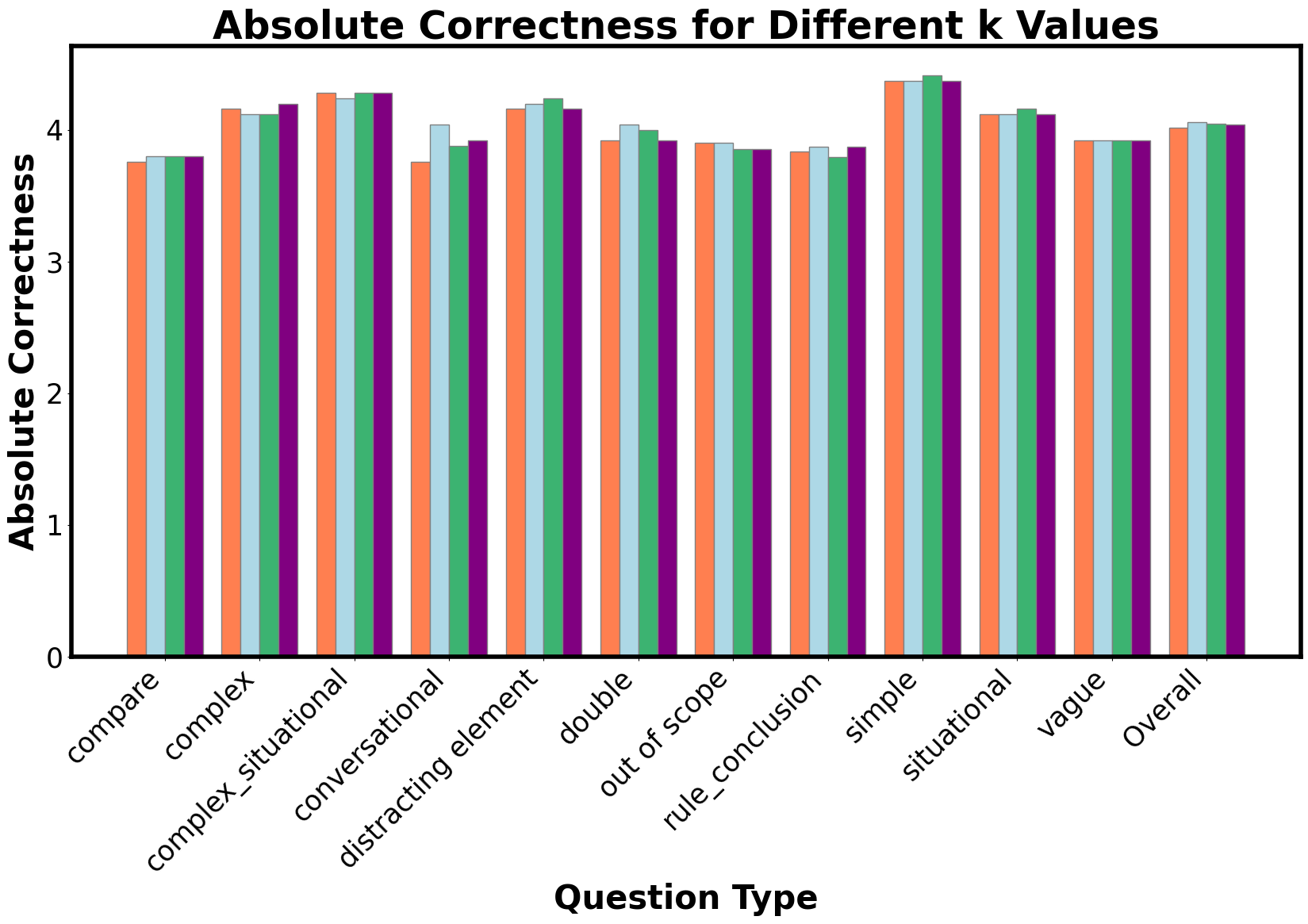}
  \caption {RAG Evaluation Metrics for Varied Top-$k$} \label{fig:eval-metrics-vary-topk}
\end{figure*}

\FloatBarrier
\subsection{Human Annotation Criteria}
\label{subsec:annotation-criteria}

\begin{table*}[h]
  \centering
  \begin{tabular}{p{1cm}p{3cm}p{10cm}}
    \hline
    \textbf{No.} & \textbf{Criterion} & \textbf{Description} \\
    \hline
    1 & \textbf{Faithfulness} & Are all claims in the answer inferred from the context? \\
    \hline
    2 & \textbf{Answer Relevancy} & Is the answer relevant to the question? \\
    \hline
    3 & \textbf{Context Relevancy} & Is the context relevant to the question? \\
    \hline
    4 & \textbf{Correctness} & Is the answer correct, given the context? \\
    \hline
    5 & \textbf{Clarity} & Is the answer clear and free of extensive jargon? \\
    \hline
    6 & \textbf{Completeness} & Does the answer fully address all parts and sub-questions? \\
    \hline
  \end{tabular}
  \caption{\label{tab:evaluation_criteria} Criteria for evaluating the quality of QA pairs.}
\end{table*}

\FloatBarrier
\subsection{Parameter Mappings}

\subsubsection{Top-$k$ ($k$) and Number of Query Rewrites ($Q$)}
\label{appendix:param-mappings-1}

\begin{table}[h]
\centering
\resizebox{\columnwidth}{!}{%
\begin{tabular}{p{3.5cm}p{1.3cm}p{5.5cm}p{2cm}p{2cm}}
\toprule
\textbf{Parameter} & \textbf{Symbol} & \textbf{Description} & \textbf{2-Class Mappings} & \textbf{3-Class Mappings} \\ \midrule
Number of Query Rewrites & \( Q \)& Number of sub-queries generated for the original query & 0: \(Q=3\) & 0: \(Q=3\)\\
 &  &  & 1: \(Q=5\) & 1: \(Q=5\)\\
 &  &  &  & 2: \(Q=7\)\\ \midrule
Top-\( k \) Value & \( k \) & Number of top documents or contexts retrieved for processing & 0: \( k = 5 \) & 0: \( k = 3 \) \\
 &  &  & 1: \( k = 10 \) & 1: \( k = 5 \) \\
 &  &  &  & 2: \( k = 7 \) \\ \bottomrule
\end{tabular}
}
\caption{Parameter Symbols, Descriptions, and Mappings}
\label{tab:parameter_mappings_part1}
\end{table}

\FloatBarrier
\clearpage
\subsubsection{Maximum Keywords ($K$) and Maximum Sequence Length ($S$)}

\begin{table}[h]
\centering
\resizebox{\linewidth}{!}{%
\begin{tabular}{@{}p{3.5cm}p{1.3cm}p{5.5cm}p{2cm}p{2.cm}@{}}
\toprule
\textbf{Parameter} & \textbf{Symbol} & \textbf{Description} & \textbf{2-Class Mappings} & \textbf{3-Class Mappings} \\ \midrule
Max Keywords per Query & \( K \)& Maximum number of keywords used per query for KG retrieval & 0: \( K =4 \)& 0:  \( K =3 \)\\
 &  &  & 1: \( K = 5\)& 1:  \( K =4 \)\\
 &  &  &  & 2:  \( K = 5 \)\\ \midrule
Max Knowledge Sequence & \( S \)& Maximum sequence length for knowledge graph paths & 0: \( S = 2\)& 0: \( S = 1\)\\
 &  &  & 1: \( S = 3 \)& 1: \( S = 2 \)\\
 &  &  &  & 2: \( S = 3 \)\\ \bottomrule
\end{tabular}
}
\caption{Parameter Symbols, Descriptions, and Mappings (Part 2)}
\label{tab:parameter_mappings_part2}
\end{table}

\FloatBarrier
\twocolumn
\subsection{Correctness Evaluator Prompts}
\label{appendix:correctness-eval-prompts}
\subsubsection{Method 1: LLamaIndex CorrectnessEvaluator}
\vspace{15pt}
\noindent\rule{\linewidth}{0.4pt}
\vspace{15pt}
\footnotesize
You are an expert evaluation system for a question answering chatbot. You are given the following information:
\begin{itemize}[leftmargin=*]
    \item a user query,
    \item a reference answer, and
    \item a generated answer.
\end{itemize}

Your job is to judge the relevance and correctness of the generated answer. Output a single score that represents a holistic evaluation. You must return your response in a line with only the score. Do not return answers in any other format. On a separate line, provide your reasoning for the score as well.

Follow these guidelines for scoring:
\begin{itemize}[leftmargin=*]
    \item Your score has to be between 1 and 5, where 1 is the worst and 5 is the best.
    \item If the generated answer is not relevant to the user query, give a score of 1.
    \item If the generated answer is relevant but contains mistakes, give a score between 2 and 3.
    \item If the generated answer is relevant and fully correct, give a score between 4 and 5.
\end{itemize}
\vspace{15pt}
\noindent\rule{\linewidth}{0.4pt}
\vspace{15pt}

\FloatBarrier
\subsubsection{Method 2: Custom Prompt 1}
\vspace{15pt}
\noindent\rule{\linewidth}{0.4pt}
\vspace{15pt}
\footnotesize
You are an expert evaluation system for a question answering chatbot. You are given the following information:
\begin{itemize}[leftmargin=*]
    \item a user query,
    \item a reference answer, and
    \item a generated answer.
\end{itemize}

Your job is to judge the correctness of the generated answer. Output a single score that represents a holistic evaluation. You must return your response in a line with only the score. Do not return answers in any other format. On a separate line, provide your reasoning for the score as well.

Follow these guidelines for scoring:
\begin{itemize}[leftmargin=*]
    \item Your score has to be between 1 and 5, where 1 is the worst and 5 is the best.
    \item Use the following criteria for scoring correctness:
\end{itemize}

\begin{itemize}[leftmargin=*]
    \item[1.] Score of 1:
    \begin{itemize}[leftmargin=*]
        \item The generated answer is completely incorrect.
        \item Contains major factual errors or misconceptions.
        \item Does not address any components of the user query correctly.
    \end{itemize}
    \item[2.] Score of 2:
    \begin{itemize}[leftmargin=*]
        \item The generated answer has significant mistakes.
        \item Addresses at least one component of the user query correctly but has major errors in other parts.
    \end{itemize}
    \item[3.] Score of 3:
    \begin{itemize}[leftmargin=*]
        \item The generated answer is partially correct.
        \item Addresses multiple components of the user query correctly but includes some incorrect information.
        \item Minor factual errors are present.
    \end{itemize}
    \item[4.] Score of 4:
    \begin{itemize}[leftmargin=*]
        \item The generated answer is mostly correct.
        \item Correctly addresses all components of the user query with minimal errors.
        \item Errors do not substantially affect the overall correctness.
    \end{itemize}
    \item[5.] Score of 5:
    \begin{itemize}[leftmargin=*]
        \item The generated answer is completely correct.
        \item Addresses all components of the user query correctly without any errors.
        \item The answer is factually accurate and aligns perfectly with the reference answer.
    \end{itemize}
\end{itemize}
\vspace{15pt}
\noindent\rule{\linewidth}{0.4pt}
\vspace{15pt}

\FloatBarrier
\subsubsection{Method 3: Custom Prompt 2}
\vspace{15pt}
\noindent\rule{\linewidth}{0.4pt}
\vspace{15pt}
\footnotesize
You are an expert evaluation system for a question answering chatbot. You are given the following information:
\begin{itemize}[leftmargin=*]
    \item a user query,
    \item a reference answer, and
    \item a generated answer.
\end{itemize}

Your job is to judge the correctness of the generated answer. Output a single score that represents a holistic evaluation. You must return your response in a line with only the score. Do not return answers in any other format. On a separate line, provide your reasoning for the score as well. The reasoning must not exceed one sentence.

Follow these guidelines for scoring:
\begin{itemize}[leftmargin=*]
    \item Your score has to be between 1 and 5, where 1 is the worst and 5 is the best.
    \item Use the following criteria for scoring correctness:
\end{itemize}

\begin{itemize}[leftmargin=*]
    \item[1.] Score of 1:
    \begin{itemize}[leftmargin=*]
        \item The generated answer is completely incorrect.
        \item Contains major factual errors or misconceptions.
        \item Does not address any components of the user query correctly.
        \item Example: \\
            Query: "What is the capital of France?" \\
            Generated Answer: "The capital of France is Berlin."
    \end{itemize}
    \item[2.] Score of 2:
    \begin{itemize}[leftmargin=*]
        \item Significant mistakes are present.
        \item Addresses at least one component of the user query correctly but has major errors in other parts.
        \item Example: \\
            Query: "What is the capital of France and its population?" \\
            Generated Answer: "The capital of France is Paris, and its population is 100 million."
    \end{itemize}
    \item[3.] Score of 3:
    \begin{itemize}[leftmargin=*]
        \item Partially correct with some incorrect information.
        \item Addresses multiple components of the user query correctly.
        \item Minor factual errors are present.
        \item Example: \\
            Query: "What is the capital of France and its population?" \\
            Generated Answer: "The capital of France is Paris, and its population is around 3 million."
    \end{itemize}
    \item[4.] Score of 4:
    \begin{itemize}[leftmargin=*]
        \item Mostly correct with minimal errors.
        \item Correctly addresses all components of the user query.
        \item Errors do not substantially affect the overall correctness.
        \item Example: \\
            Query: "What is the capital of France and its population?" \\
            Generated Answer: "The capital of France is Paris, and its population is approximately 2.1 million."
    \end{itemize}
    \item[5.] Score of 5:
    \begin{itemize}[leftmargin=*]
        \item Completely correct.
        \item Addresses all components of the user query correctly without any errors.
        \item Providing more information than necessary should not be penalized as long as all provided information is correct.
        \item Example: \\
            Query: "What is the capital of France and its population?" \\
            Generated Answer: "The capital of France is Paris, and its population is approximately 2.1 million. Paris is known for its rich history and iconic landmarks such as the Eiffel Tower and Notre-Dame Cathedral."
    \end{itemize}
\end{itemize}

Checklist for Evaluation:
\begin{itemize}[leftmargin=*]
    \item Component Coverage: Does the answer cover all parts of the query?
    \item Factual Accuracy: Are the facts presented in the answer correct?
    \item Error Severity: How severe are any errors present in the answer?
    \item Comparison to Reference: How closely does the answer align with the reference answer?
\end{itemize}

Edge Cases:
\begin{itemize}[leftmargin=*]
    \item If the answer includes both correct and completely irrelevant information, focus only on the relevant portions for scoring.
    \item If the answer is correct but incomplete, score based on the completeness criteria within the relevant score range.
    \item If the answer provides more information than necessary, it should not be penalized as long as all information is correct.
\end{itemize}

\vspace{15pt}
\noindent\rule{\linewidth}{0.4pt}
\vspace{15pt}

\FloatBarrier
\onecolumn
\subsection{Correctness Evaluator Results}
\label{appendix:correctness-eval-results}

\begin{figure*}[h]
  \includegraphics[width=0.48\linewidth]{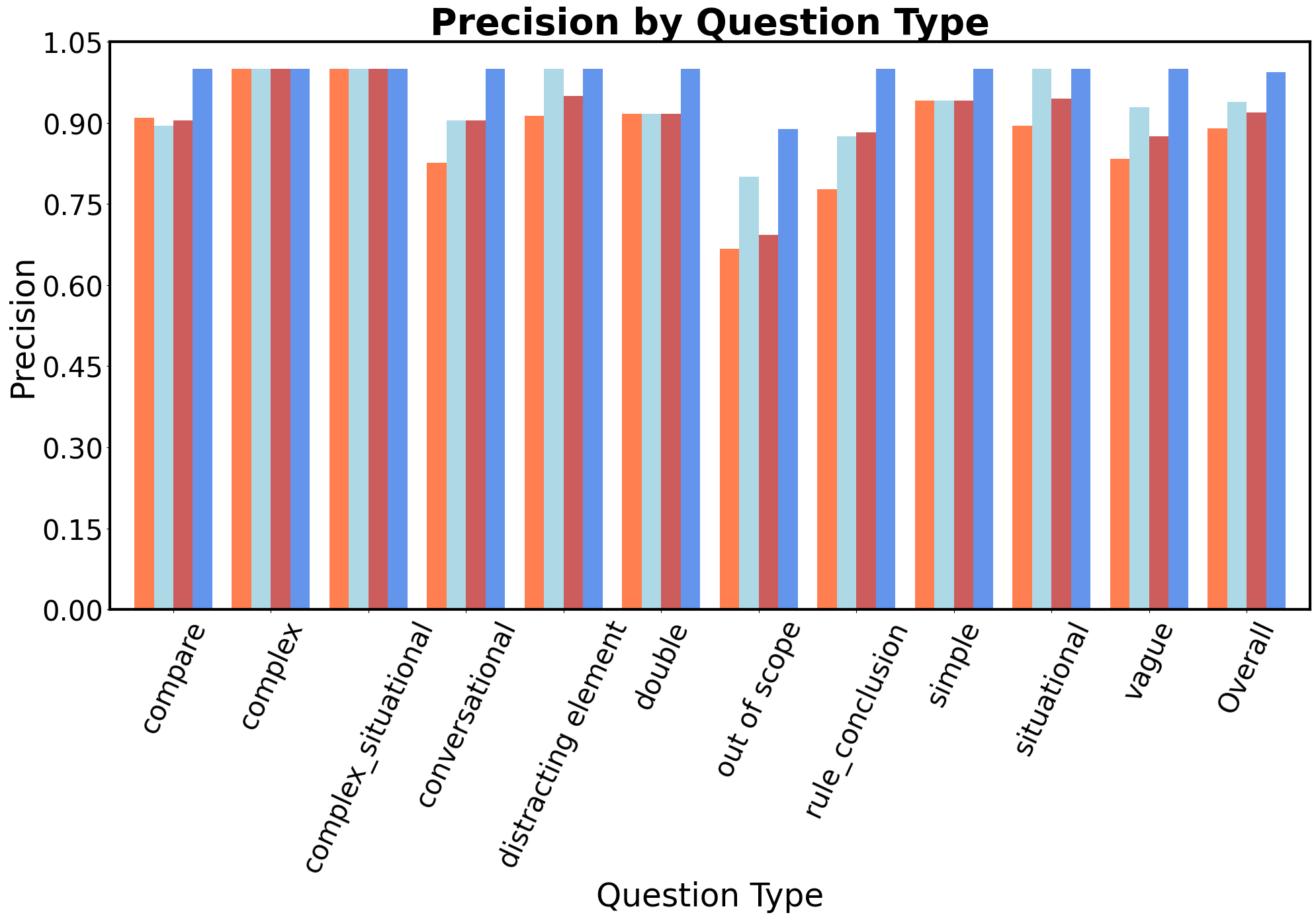} \hfill
  \includegraphics[width=0.48\linewidth]{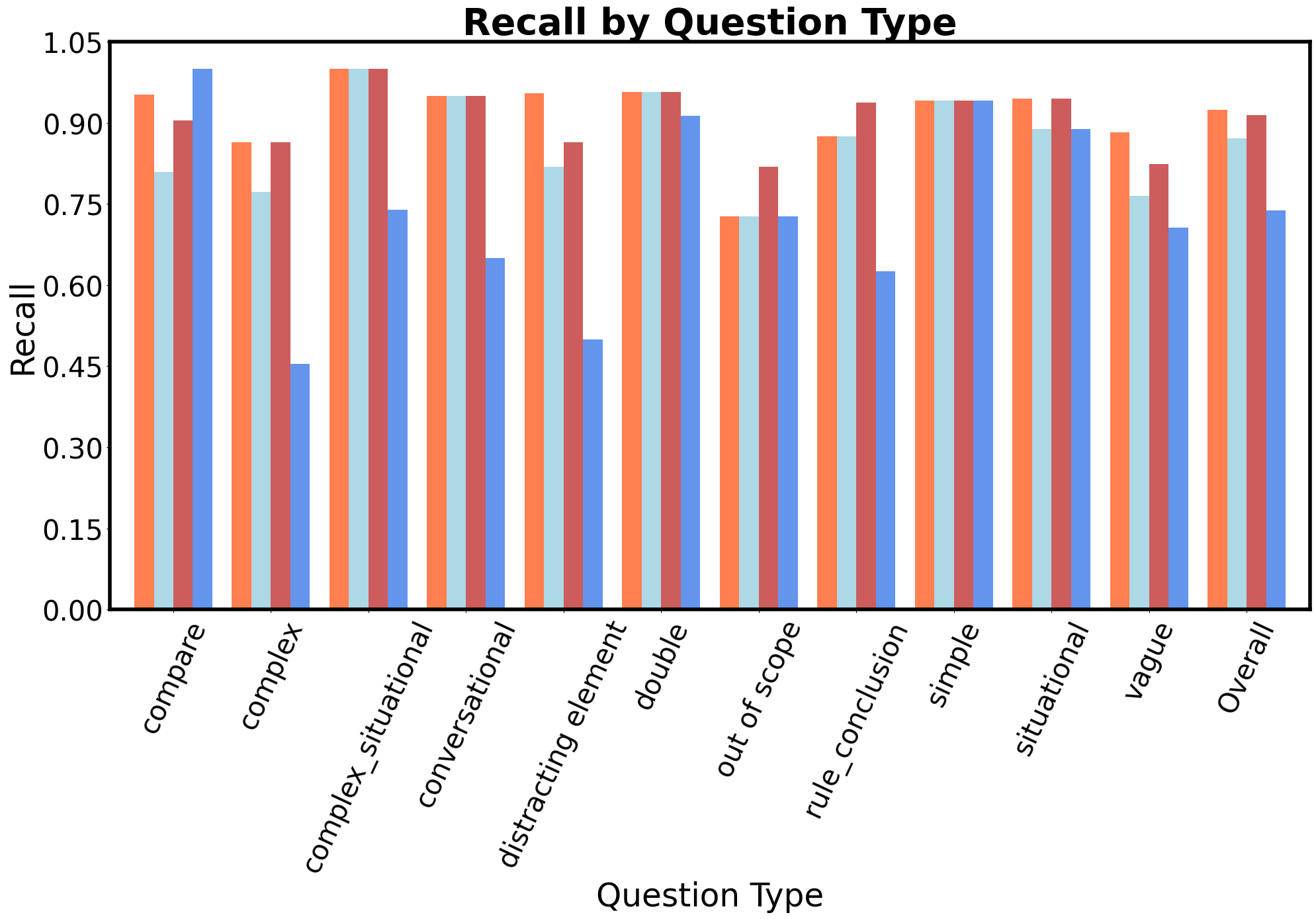}
  \includegraphics[width=0.48\linewidth]{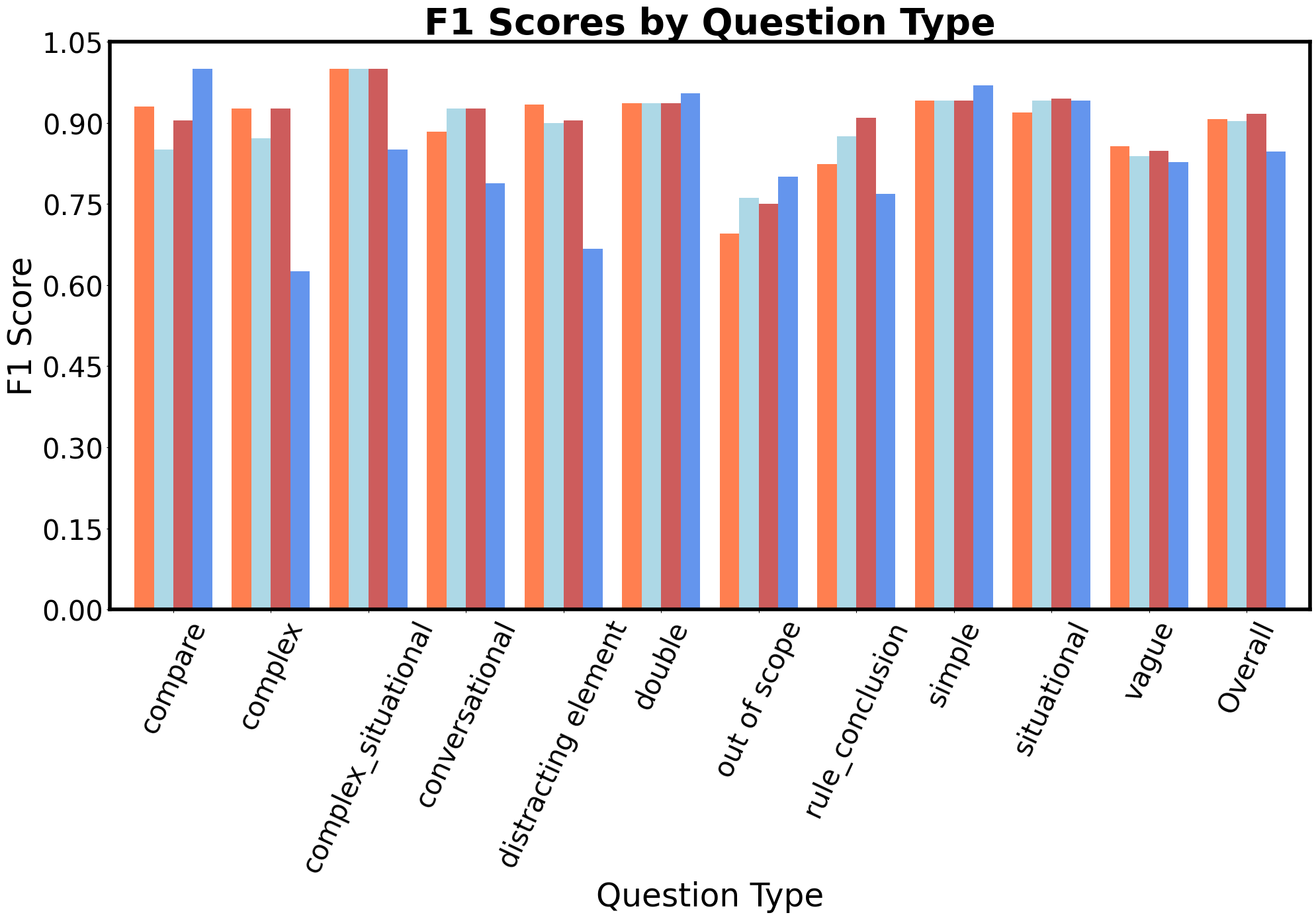} \hspace{15pt}
  \includegraphics[width=0.48\linewidth]{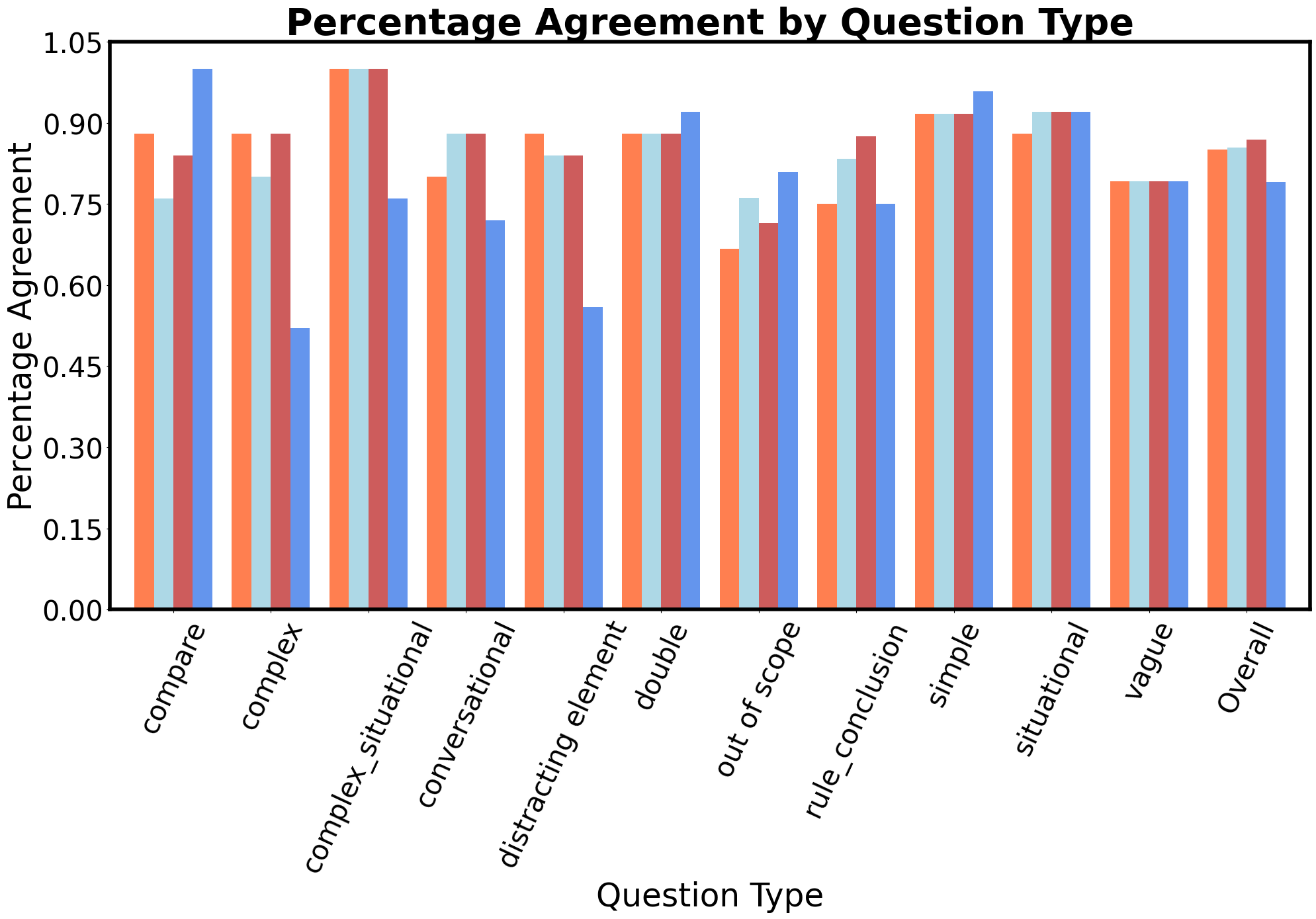}
  \caption {Precision, recall, F1 score, and percentage agreement of the prompt-based (1-5 scale) LLM-as-a-judge correctness evaluation compared to human judgments.}
  \label{fig:correctness_classification_metrics}
\end{figure*}

\begin{figure*}[h]
  \includegraphics[width=0.48\linewidth]{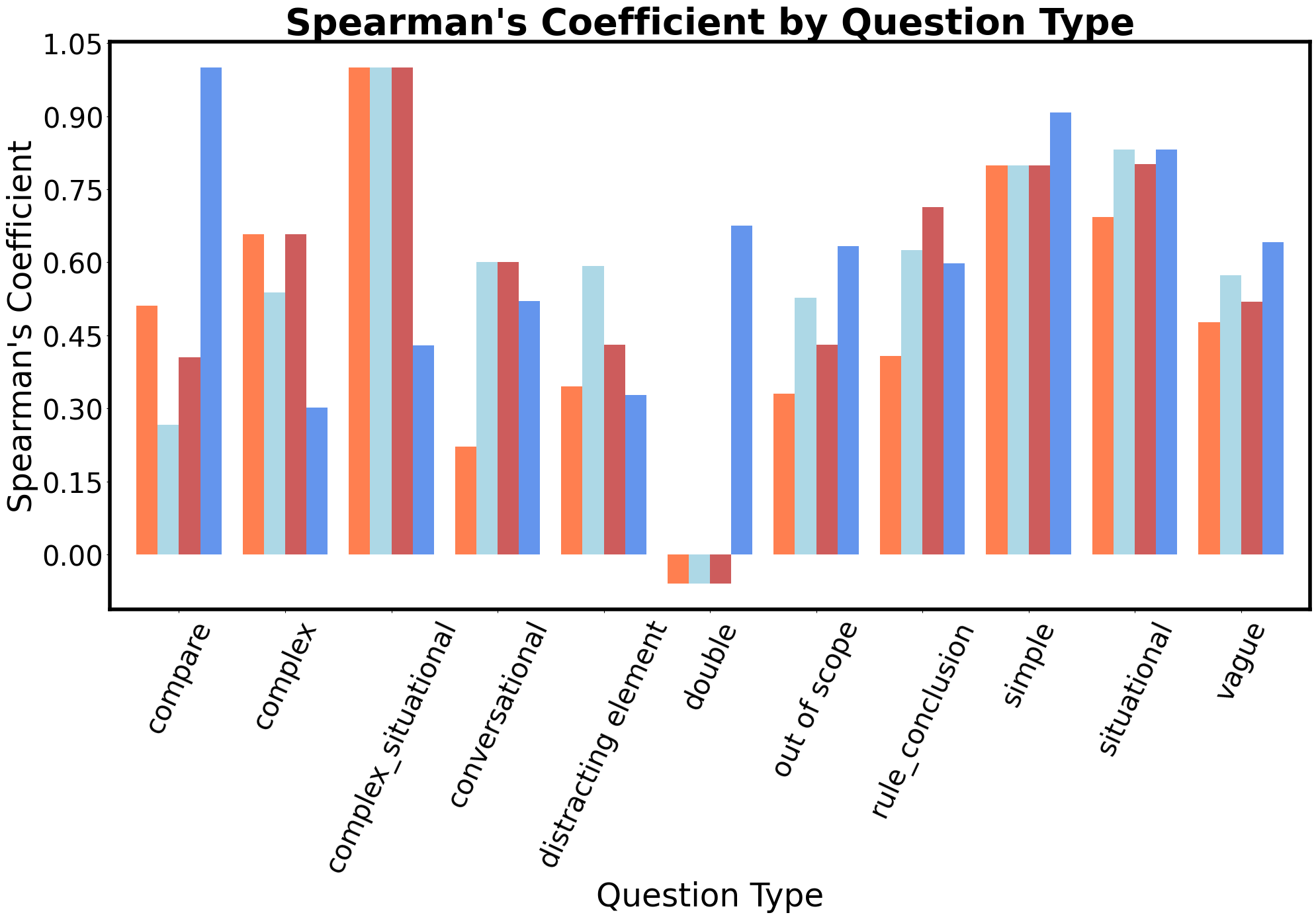} \hfill
  \includegraphics[width=0.48\linewidth]{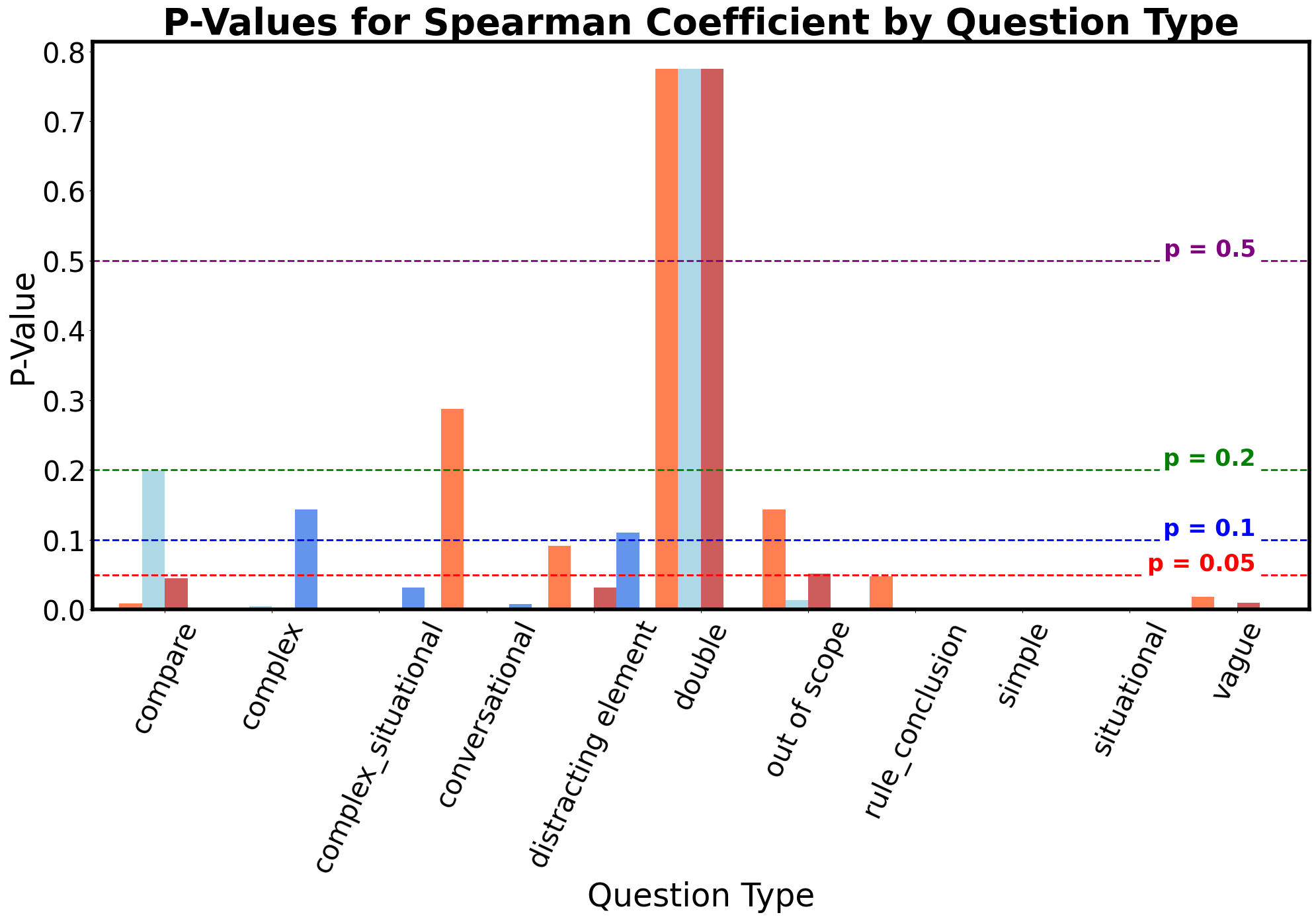}
  \caption {Spearman Coefficient comparing our custom LLM-as-a-judge (1-5 scale) prompts with Giskard's binary correctness evaluator for each question type. The second plot displays the p-values.}
  \label{fig:correctness_spearman_metrics}
\end{figure*}

\FloatBarrier
\clearpage
\subsection{Classifier Data Augmentation Prompts}
\label{appendix:classifier-data-aug-prompts}
\subsubsection{Vague Prompt}
Rewrite the following question to be more vague, but it must still require the same number of pieces of information to answer. For example, a definition is one piece of information. A definition and an explanation of the concept are two separate pieces of information. Do not add or remove any pieces of information, and do not alter the fundamental meaning of the question. Output only the rewritten question, absolutely nothing else: \verb|{question}|

\subsubsection{Verbose Prompt}
Rewrite the following question to be more verbose, but it must still require the same number of pieces of information to answer. For example, a definition is one piece of information. A definition and an explanation of the concept are two separate pieces of information. Do not add or remove any pieces of information, and do not alter the fundamental meaning of the question. Output only the rewritten question, absolutely nothing else: \verb|{question}|

\subsubsection{Concise Prompt}
Rewrite the following question to be more concise, but it must still require the same number of pieces of information to answer. For example, a definition is one piece of information. A definition and an explanation of the concept are two separate pieces of information. Do not add or remove any pieces of information, and do not alter the fundamental meaning of the question. Output only the rewritten question, absolutely nothing else: \verb|{question}|

\FloatBarrier
\subsection{2-Class Classifier Results}
\label{appendix:2-class-classifier-results}

\begin{table}[h]
  \centering
  \resizebox{\columnwidth}{!}{%
  \begin{tabular}{lccc}
    \hline
    \textbf{Model} & \textbf{Precision} & \textbf{Recall} & \textbf{F1 Score} \\
    \hline
    Random Labels & 0.49 & 0.49 & 0.49 \\
    facebook/bart-large-mnli & 0.55 & 0.55 & 0.53 \\
    DeBERTa-v3-base-mnli-fever-anli & 0.59 & 0.57 & 0.56 \\
    Logistic Regression (TF-IDF) & 0.88 & 0.88 & 0.88 \\
    SVM (TF-IDF) & 0.92 & 0.92 & 0.92 \\
    distilbert-base-uncased finetuned & 0.92 & 0.92 & 0.92 \\
    \hline
  \end{tabular}
  }
  \caption{2-Class Classification Results}
  \label{tab:classification-results-2class}
\end{table}

\FloatBarrier

\subsection{3-Class Ablation Results}
\label{appendix:3-class-ablation}

\begin{table*}[!h]
  \centering
  \begin{tabular}{p{3.5cm}p{2.5cm}p{2.5cm}p{3.cm}p{2.5cm}}
    \hline
    \textbf{Method} & \textbf{Faithfulness} & \textbf{Answer \newline Relevancy} & \textbf{Absolute \newline Correctness (1-5)} & \textbf{Correctness \newline (Threshold=4.0)} \\
    \hline
    $k$ & 0.7723 & \textbf{0.7940} & 4.0409 & 0.7621 \\
    $k$, $Q$ & \underline{0.8971} & 0.7778 & \textbf{4.2528} & 0.8141 \\
    $k$, $Q$ + reranker & \textbf{0.9098} & \underline{0.7902} & \underline{4.2342} & \underline{0.8178} \\
    $k$, $K^*$, $S^*$ & 0.8733 & 0.7635 & 4.1227 & 0.8141 \\
    $k$, $K$, $S$ & 0.8660 & 0.7780 & 4.1822 & 0.8030 \\
    $k$, $K$, $S$ + reranker & 0.8821 & 0.7872 & 4.1858 & \underline{0.8178} \\
    $k$, $K$, $S$, $Q$ & 0.8465 & 0.7734 & 4.1338 & 0.7918 \\
    $k$, $K$, $S$, $Q$ + reranker & 0.8689 & 0.7853 & 4.1859 & \textbf{0.8402} \\
    \hline
  \end{tabular}
  \caption{\label{tab:ablation-3-class} Ablation study results for different configurations of adaptive $k$ in a 3-class setting. For descriptions of parameters, refer to Table \ref{tab:combined-results-metrics}. The highest value in each column is highlighted in bold, and the second highest value is underlined. The * indicates parameters held fixed, rather than adaptive.}
\end{table*}

\subsection{2-Class Ablation Results}
\label{appendix:2-class-ablation}

\begin{table*}[!h]
  \centering
  \begin{tabular}{p{4cm}p{2.5cm}p{2.5cm}p{2.5cm}p{2.5cm}}
    \hline
    \textbf{Method} & \textbf{Faithfulness} & \textbf{Answer \newline Relevancy} & \textbf{Absolute \newline Correctness \newline (1-5)} & \textbf{Correctness \newline (Threshold=4.0)} \\
    \hline
    $k$ & 0.8111 & 0.7835 & 4.0372 & 0.7546 \\
    $k$, $K^*$, $S^*$ & 0.8725 & \underline{0.7830} & 4.1115 & \underline{0.8216} \\
    $k$, $K$, $S$ & 0.8551 & 0.7810 & 4.1487 & 0.7955 \\
    $k$, $K$, $S$ + reranker & \textbf{0.8792} & \textbf{0.7878} & \textbf{4.1710} & 0.8141 \\
    $k$, $K$, $S$ + adaptive $Q$ & 0.8328 & 0.7800 & 4.0558 & 0.7770 \\
     $k$, $K$, $S$ + $Q$ + reranker & \underline{0.8765} & 0.7803 & \underline{4.1636} & \textbf{0.8253} \\
    \hline
  \end{tabular}
  \caption{\label{tab:ablation-2-class} Ablation study results for different configurations starting from adaptive $k$. The highest value in each column is highlighted in bold, and the second highest value is underlined.}
\end{table*}

\subsection{Future Work and Limitations}
\label{appendix:future-work-limitations}

This study has several limitations that suggest areas for future improvement. Correctness evaluation is limited by reliance on a single evaluator familiar with the policy corpus. Averaging a larger quantity of human evaluations would improve reliability. Additionally, our knowledge graph construction process may be improved. For instance, using LLM-based methods for de-duplication and/or custom Cypher query generation to improve context retrieval and precision. Furthermore, our parameter mappings were not rigourously validated quantitatively. Further evaluation of parameter selections could provide better mappings as well as upper and lower bounds to performance. The classifier was trained using domain-specific synthetically generated data - which, though we inject significant noise, may harbour the LLM's own unconcious biases in terms of structure - possibly limiting the generalisability of the classifier on unseen user queries. Also, more classification categories e.g. 4 or 5-class, would permit more granular parameter selections and potentially greater efficiency improvements. Another limitation is that while LL144 is included in the GPT models' training data, subsequent minor revisions may affect the accuracy of these baseline methods.


Integrating human feedback into the evaluation loop could better align metrics with user preferences and validate performance metrics in real-world settings. Future work should also consider fine-tuning the LLM using techniques like RLHF \citep{Bai2022TrainingAH}, RLAIF \citep{Lee2023RLAIFSR}, or other preference optimisation methods \citep{Song2023PreferenceRO}. Further, refining the query rewriter \citep{Ma2023QueryRF, Mao2024RaFeRF} and exploring iterative answer refinement \citep{asai2023selfrag} could enhance metrics like relevancy and correctness.

\end{document}